\documentclass[a4paper,aps,pra,reprint,superscriptaddress, floatfix,twocolumn]{revtex4-1}
\usepackage{amsmath}
\usepackage{verbatim}
\usepackage{graphicx}
\usepackage{physics}
\usepackage{color}
\usepackage[normalem]{ulem}
\usepackage{amsmath,amsfonts,amssymb,graphicx,color,times,bbm,psfrag,dsfont,slashed,bm}
\usepackage[colorlinks=true,citecolor=blue,linkcolor=blue,urlcolor=blue]{hyperref}
\usepackage[usenames,dvipsnames]{xcolor}
\usepackage{gensymb}
\usepackage{bbold}
\usepackage{float}
\usepackage[T1]{fontenc}
\usepackage[latin9]{inputenc}

\begin{document}

\preprint{}

\title{Stark localization near Aubry-Andr{\'e} criticality}

\author{Ayan Sahoo}
\affiliation{Harish-Chandra Research Institute, A CI of Homi Bhabha National Institute, Allahabad 211019, India}

\author{Aitijhya Saha}\affiliation{Harish-Chandra Research Institute, A CI of Homi Bhabha National Institute, Allahabad 211019, India}\affiliation{Department of Theoretical Physics, Tata Institute of Fundamental Research, Mumbai 400005, India}

\author{Debraj Rakshit}
\affiliation{Harish-Chandra Research Institute, A CI of Homi Bhabha National Institute, Allahabad 211019, India}


\begin{abstract}
 In this work, we investigate the Stark localization near the Aubry-Andr\'{e} (AA) critical point.  We perform careful studies for reporting system-dependent parameters, such as localization length, inverse participation ratio (IPR), and energy gap between the ground and first excited state, for characterizing the localization-delocalization transition. We show that the scaling exponents possessed by these key descriptors of localization are quite different from that of a pure AA model or Stark model. Near the critical point of the AA model, in the presence of Stark field of strength $h$, the localization length $\zeta$ scales as $\zeta\propto h^{-\nu}$ with $\nu\approx0.29$  which is different than both the pure AA model ($\nu=1$) and Stark model ($\nu\approx0.33$). The IPR in this case scales as IPR $\propto h^{s}$ with $s\approx0.096$ which is again significantly different than both the pure AA model ($s\approx0.33$) and Stark model ($s\approx0.33$). The energy gap, $\Delta$, scales as $E\propto h^{\nu z}$, where $z\approx2.37$ which is however same as the pure AA model. Finally, we discuss how invoking a criticality inducing additional control parameter may help in designing better many-body quantum sensors. Quantum critical sensors exploit the venerability of the wavefunction near the quantum critical point against small parameter shifts. By incorporating a control parameter in the form of the quasi-periodic field, i.e., the AA potential, we show a significant advantage can be drawn in estimating an unknown parameter, which is considered here to be the Stark weak field strength, with high precision. We compute and extract the scaling exponent associated with the Quantum Fisher Information (QFI), $F_Q$, which provides a bound on the uncertainty associated with measuring an unknown parameter. We demonstrate that by a control field in the form of the quasiperiodic potential, and setting it at AA criticality, $F_Q$ scales with the system size as, $F_{Q}\propto L^{\beta}$, where $\beta \approx 6.7$, which is greater in comparison to the pure Stark case, for which $\beta$ is known to be $5.8$. We, thus, pitch a generic idea of invoking quantum criticality inducing additional control parameters for obtaining enhanced quantum advantage in many-body critical sensors.
\end{abstract}

\maketitle

\section{Introduction} \textcolor{black}{Randomness is an inherent property of nature whose influence on any physical system often defies our intuition and yields non-trivial effects.} Consequently, disordered systems have long been a focal point of significant research endeavors~\cite{Binder86,Belitz05,Das08,Alloul09,Konar22,Ghosh21,Aitijhaya23,roy20}. Within the realm of quantum mechanics, the interplay of disorder with fundamental principles gives rise to a myriad of noteworthy phenomena, including the emergence of novel quantum phases, order-from-disorder phenomena~\cite{Aizenman89,Wehr06,Bera14,Bera16,Bera17,Mishra16,Sadhukhan15,Sadhukhan16}, the enigmatic high-$T_c$ superconductivity~\cite{Auerbach94}, and the phenomenon of localization \cite{Anderson58,Abrahams79,Evers08,Nandkishore15,Abanin19}. Among these phenomena, disorder-induced Anderson localization occupies a prominent position \cite{Yamilov23,Lagendijk09}. This well-established concept elucidates how the presence of randomness can hinder the diffusion of waves within a system, showcasing its pivotal role in modern condensed matter physics. \textcolor{black}{In Anderson's model the disorder is truly random and uncorrelated. Whereas some disorders are not random, rather there is some correlation. Quasi-periodicity is the candidate that qualifies for this correlated disorder.}

\textcolor{black}{In the recent past and even nowadays quasiperiodic systems \cite{Sinha19,Wei19} are widely studied as it captures very exotic features of various novel phases of matter and supports phase transition between those states \cite{Chowdhury86,Mezard87,Sachdev99,Yao14,Zuniga13}. Among these models, the Aubry-Andr\'{e} (AA)~\cite{Aubry80,Sahoo24} model is of central interest of researchers for its distinct characteristics in the realm of phase transitions \cite{Zanardi2006,Mondal23}.} One of the distinguishing features of the AA model is its self-dual symmetry, evident between the Hamiltonian formulations in momentum and position spaces. This symmetry manifests in an energy-independent localization-delocalization transition~\cite{Michal14, Iyer13} occurring at a finite modulation strength. This model has been extensively explored across various contexts, including the elucidation of the Hofstadter butterfly structure of the energy spectrum under controlled parameters~\cite{Hofstadter76}, investigations into transport phenomena~\cite{Purkayastha18,Sutradhar19}, analysis of mobility edges~\cite{Saha16,Ganeshan15,Ganeshan1502}, critical behavior, and exploration of topological phases~\cite{DeGottardi13,Cai13,Fraxanet3}, many-body localization \cite{Gu14,Zhang08,Zhang09}, dynamical phase transition \cite{Modak21},  in context of open quantum system \cite{Carmichael93,Xu16,Kawabata17,Lee19}, and for its applications in quantum technology \cite{Sahoo24}. \textcolor{black}{The AA model has been realized using ultracold atoms in an incommensurate optical lattice \cite{Makris08,Klaiman08,Ruter10} and photonic lattices \cite{Roati08}.}

\textcolor{black} {The AA model is a tight binding model with nearest neighbor tunneling in the presence of sinusoidal quasi-periodic potential. When the strength of the potential becomes twice the tunneling parameter, in the thermodynamic limit, the system experiences a phase transition from a delocalized phase to a localized phase. On the other hand, in a tight binding model, if a gradient field is induced across the lattice making the on-site energies off-resonant, the tunneling rate gets suppressed resulting in the localization of wavefunction in space in the absence of disorder. In the thermodynamic limit, this localization, known as Stark localization \cite{Schulz19}, takes place at zero field limit.} A substantial amount of research has been conducted for understanding Stark localization phenomena in context of single-particle \cite{Wannier60,Fukuyama73,Holthaus95,Kolovsky03,Kolovsky08,Kolovsky13,Nieuwenburg19} and many-body \cite{Jiang23,Zhang21,Yao21,Wei22,Doggen21} systems, and for finding applications \cite{Morong21,Preiss15,Karamlou22,He23,Kohlert21} by exploiting it.

\begin{figure}[t]
\centering
\includegraphics[width=0.42\textwidth]{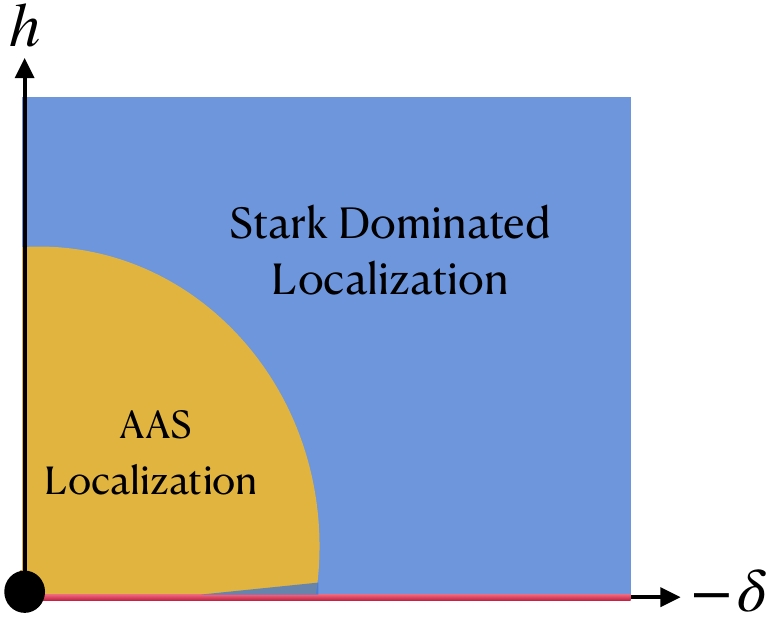}
 \caption{A schematic figure depicting the nature of localization in the presence of the dual potentials. In the thermodynamic limit, the system is undet the influence of the joint criticalities at $h \to 0$ and $\delta=0$ due to the Stark potential and the AA potential, respectively (black circle).  The system is delocalized for $|\delta| > 0$ and $h = 0$ (red line). It enters into the localized phase in the limit $h \to 0$, irrespective of the value $\delta$. However, the finite value of $\delta$ may significantly influence the nature of localization when compared to the pure Stark case, and it is then termed as AAS critical localization (yellow region). For large values of $h$ and away from  $\delta=0$, the AA potential has negligible influence on localization, and it becomes  Stark dominated, as one may expect. In the $h \to 0$ limit, this can by shown by comparing the scaling exponent associated with the localization transition with that of the pure Stark model. For non-zero $h$, we monitor the fidelity of the ground state wavefunction with the system under joint potentials with the pure Stark case.  The fidelity is expected to be high in the Stark dominated localization (blue region). }
\label{fig:schamatic}
\end{figure}

All these are quantum phase transitions that manifest at zero temperature when the control parameter of a quantum system's Hamiltonian is finely tuned to a critical value, referred to as the quantum critical point (QCP). At this juncture, the system exhibits scaling and universality~\cite{Hikami81,Altland97,Wang21,Luo21,LuoR22,Luo22,Bu22}, meaning that the equilibrium properties of certain physical observables near the QCP can be succinctly described by a handful of critical exponents. Extracting these critical exponents necessitates an understanding of various physical quantities, such as the localization length, Inverse Participation Ratio (IPR),  energy gap, fidelity susceptibility, etc. These quantities provide essential insights into the behavior of the system near the QCP and facilitate the characterization of its critical behavior. In this work, we consider a single particle tight-binding model with nearest-neighbor hopping. Particularly, quantum criticality in the disordered AA model has been studied recently and a hybrid scaling theory has been developed \cite{Bu22}. In a similar spirit, we ask the following question: What is the nature of the quantum criticality of the AA system in the presence of an additional Stark potential? The Stark potential is always localizing. However, the interplay of these potentials ensures that the system has a rich quantum of critical behavior. The system is first prepared at the QCP of the AA model. Then a Stark potential is introduced for studying the scaling properties of localization length, IPR, and energy gap between the ground and the first excited state. We perform several case studies that include scaling studies when the AAH field is set (i) right at the AAH criticality, (ii) near the AAH criticality, (iii) far from the AAH criticality, and (iv) a case, where the system size, $L$, is kept fixed and different AAH field strengths are considered in order a perform a hybrid scaling analysis. The modified nature of the criticality of the system under the joint presence of the localizing potentials is completely captured in the associated scaling exponents. We perform a careful error analysis for reporting the critical exponents obtained from the scaling analysis. The scaling exponents are estimated to have uncertainties of the order of 1\%. It's worth mentioning that the scaling theory in the AAS model can be formulated for the range of the AA potential that corresponds to the delocalized phase up to the AA criticality in a pure AA model. Beyond the AA criticality where the  AA potential induces a finite localization in the pure AA model, any kind of scaling analysis in the AAS model fails. One may understand that for the regime of the AA potential, where scaling theory can be properly devised, the scaling exponents in the AAS system is strongly modified when the AA potential is at or near the AA criticality. It will be shown that far away from the AA criticality, the AA potential exerts negligible influence and the physics is dominated by the Stark potential. As a result, the scaling exponent describing the criticality of the AAS system becomes identical to the pure Stark system. Then the ground state wave function of the AAS system has high fidelity with the ground state of the pure Stark model. This fidelity based analysis is extended for identifying the parametric regime where localization properties are basically dominated by the Stark potential alone.

 Apart from the fundamental interest, we show that the system under study has important applications in quantum metrology \cite{QM006}. Recently, there are a lot of efforts towards designing critical quantum sensors - a new kind of many-body quantum sensing device. Several of these works have recognized quantum phase transitions,  such as first- and second-order quantum phase transition, and localization transition, as useful quantum resources that can be exploited for engineering the critical quantum sensors \cite{He23, Rams18, Sahoo24, Sahoo24stark, Mondal24}. Estimating the strength of an unknown weak field with high precision is a formidable task. A recent work has proposed a weak-field sensing device by employing Stark weak field inducing localization \cite{He23}. Moreover, the authors of this work also demonstrated in a parallel work that localization-delocation transition in the AA can be an excellent resource for designing many-body quantum sensors that can estimate an unknown field with high accuracy \cite{Sahoo24}.  Performance of these quantum sensing devices are marked via computation of Quantum Fisher Information (QFI), $F_Q$, that provides a bound in uncertainty associated with the measurement of the unknown parameter. It is known that the QFI of a quantum critical sensor is controlled by the scaling exponent associated with the correlation length (in case of second-order quantum phase transition) or localization length (in case of localization transition) \cite{He23,Sahoo24}. Now, given the fact that the interplay of these Stark and AA potentials modifies quantum critical behavior and the scaling properties of localization length, a natural question arises if it is possible to design a better Stark weak field sensor by invoking an additional control parameter in the form of the AA potential, and, in particular, by setting it near the AA criticality. The answer is affirmative. We present an analysis in this work in support of this claim.

Following an introductory overview in Section I, Section II presents the model of our system. In Section III, defining the characteristic quantities relevant to the localization-delocalization transition, the behavior of the corresponding scaling exponents at the criticality of the AA model is studied via the cost function approach. In Section IV the same scaling exponents are computed in the vicinity of the critical point of the AA model. In Section V, we explore the evolution of the scaling exponent of IPR as we move away from the critical region of the AA model and additionally, we discuss how the corresponding structures of the wave function change. Then in Section VI, we introduce a hybrid scaling exponent that is related to the other scaling exponents of the pure AA and Stark model. Finally, we discuss how the model under study can be utilized for designing better quantum sensors for precise estimation of the Stark weak field amplitude in Section VII. A conclusion is given in Section VIII.

\section{Model} The Hamiltonian of the Aubry-Andr\'{e} model with a linear gradient field across the lattice, is
\begin{eqnarray}
\label{eq:Ham}
 \hat{H} = &-\;J&\sum_{i}^{L-1}(\hat{c}^{\dagger}_{i}\hat{c}_{i+1}+\mathrm{h.c.}) + h \sum_{i}^{L-1} i \hat{c}^{\dagger}_{i}\hat{c}_{i}
 \nonumber \\&+&(2J+\delta) \sum_{i}^{L-1} \cos[2\pi (i \omega + \phi)] \hat{c}^{\dagger}_{i}\hat{c}_{i}
\end{eqnarray}
where, $\hat{c}^{\dagger}_{i} (\hat{c}_i)$ is the fermionic creation (annihilation) operator at the $i^{\mathrm{th}}$ site, $(2J+\delta)$ determines the amplitude of the quasiperiodic modulation that is controlled by an irrational frequency $\omega$. It is well known in the literature that proper scaling with system size $L$ appears if it is chosen from the Fibonacci series, $L=F_{n+1}$. For finite size systems, $\omega$ is approximated as $\omega = F_n/F_{n+1}$. Here $F_n$ and $F_{n+1}$ are two consecutive Fibonacci numbers. In the thermodynamic limit, $n \to \infty$, it has the property $\omega = \text{lim}_{n_{\to \infty}} F_n/F_{n+1} = (\sqrt{5}-1)/2$, which is known as the golden ratio. $\phi$ is a random number resulting in some random phase of quasiperiodic potential, drawn from a uniform distribution in $[0, 1]$, and $h$ quantifies the magnitude of the Stark field. The boundary condition used in this work is the open boundary condition (OBC). In the thermodynamic limit (i.e., $L \to  \infty$) for the absence of the second term, i.e., $h=0$, one finds that all the eigenstates are localized for $\delta>0$ and all the eigenstates are delocalized for $\delta<0$. On the other hand, without the last term i.e., $\delta = -2J$, in the thermodynamic limit, the system is localized for $h\to0$. For the rest of the work, we fix $J$ at unity, i.e., $J = 1$, for convenience.

Before we delve deeper into the details, we present a schematic diagram as a precursor of the results in Fig.~\ref{fig:schamatic} of the Aubery-Andr{\'e}-Stark (AAS) model in the parameter regime of our interest. In the following, we present the results.

\section{Scaling Analysis at the AA Criticality} We start our study of the behavior of the scaling exponents right at the AA criticality, i.e., $\delta=0$. In a finite system, at this point, the wavefunction of the particle is neither localized nor delocalized, rather there happens to be a continuous phase transition. To quantify the localized or delocalized nature of the wavefunction, we use the observable localization length, $\zeta$, defined as,
\begin{eqnarray}
\label{eq:zeta}
\zeta = \sqrt{\sum_{i}^{L}(i-i_c)^2\;p_i},
\end{eqnarray}
where $i$ denotes the lattice site, $p_i$ is the single-particle probability density at site $i$ and $i_c$ is the localization center having the expression $i_c=\sum_{i=1}^{L}ip_i$. Expanding the $n^{\mathrm{th}}$ normalised eigenstate of the system, $|\psi^n\rangle$, in terms of the single-particle computational basis, $|i\rangle$, such that $|\psi_n\rangle = \sum_i c^{(i)}_n |i \rangle$, $p_i$ is given by $p_i=|\langle i|\psi^n\rangle|^2=|c^{(i)}_n|^2$. If $g$ is the control parameter in a certain localization-delocalization transition and $g_c$ is the critical point then in the thermodynamic limit, near this criticality, the localization length scales as, 
\begin{eqnarray}
\label{eq:zeta}
\zeta \propto |g-g_c | ^{-\nu},
\end{eqnarray}
where $\nu$ is the scaling exponent. For the pure AA model, $|g-g_c|=\delta$ and $\nu=1$; for the pure Stark model $|g-g_c|=h$ and $\nu\approx0.33$. Near criticality, the localization length scales with system size as,
\begin{eqnarray}
\label{eq:zeta_Scaling_L}
\mathrm{\zeta} \propto L^{-1}.
\end{eqnarray}

\begin{figure}[t]
    \centering
\includegraphics[width=0.48\textwidth]{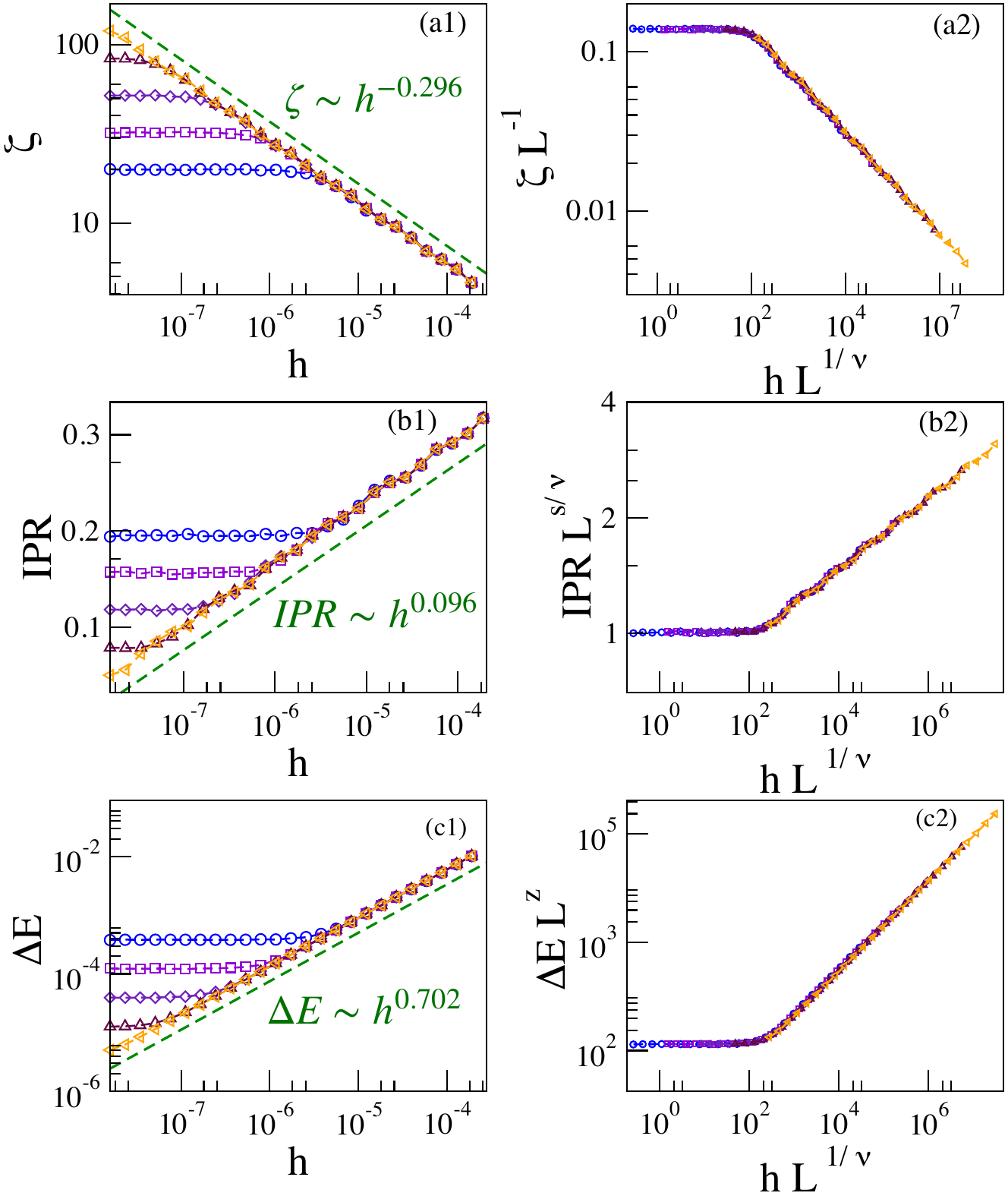}
\caption{{\bf Scaling analysis at $\delta = 0$:} {\bf (a1)} presents localization length $\zeta$ of the ground state against Stark field strength $h$ for different system sizes, $L=144$ (blue circle), $L=233$ (violet square), $L=377$ (indigo diamond), $L=610$ (maroon triangle up) and $L=987$ (orange triangle left). {\bf (a2)} shows a collapse plot for the localization length with the value of scaling exponent $\nu=0.29$. {\bf (b1)} displays the IPR versus $h$ for various values of $L$ which is aforementioned in (a1). {\bf (b2)} demonstrates the collapse plot of IPR using the value of scaling exponent $s=0.096$ and $\nu=0.29$. {\bf (c1)} shows the energy gap $\Delta E$ between the ground state and the first excited state as a function of $h$ for different $L$ values as mentioned previously. {\bf (c2)} demonstrates the collapse plot of $\Delta E$ using the scaling exponent $z=2.37$. we have taken $\phi$ average for 5000 random choices in the range $[0,1]$.}
 \label{fig:fig1}
\end{figure}

\begin{figure*}[t]
\centering
\includegraphics[width=0.95\textwidth]{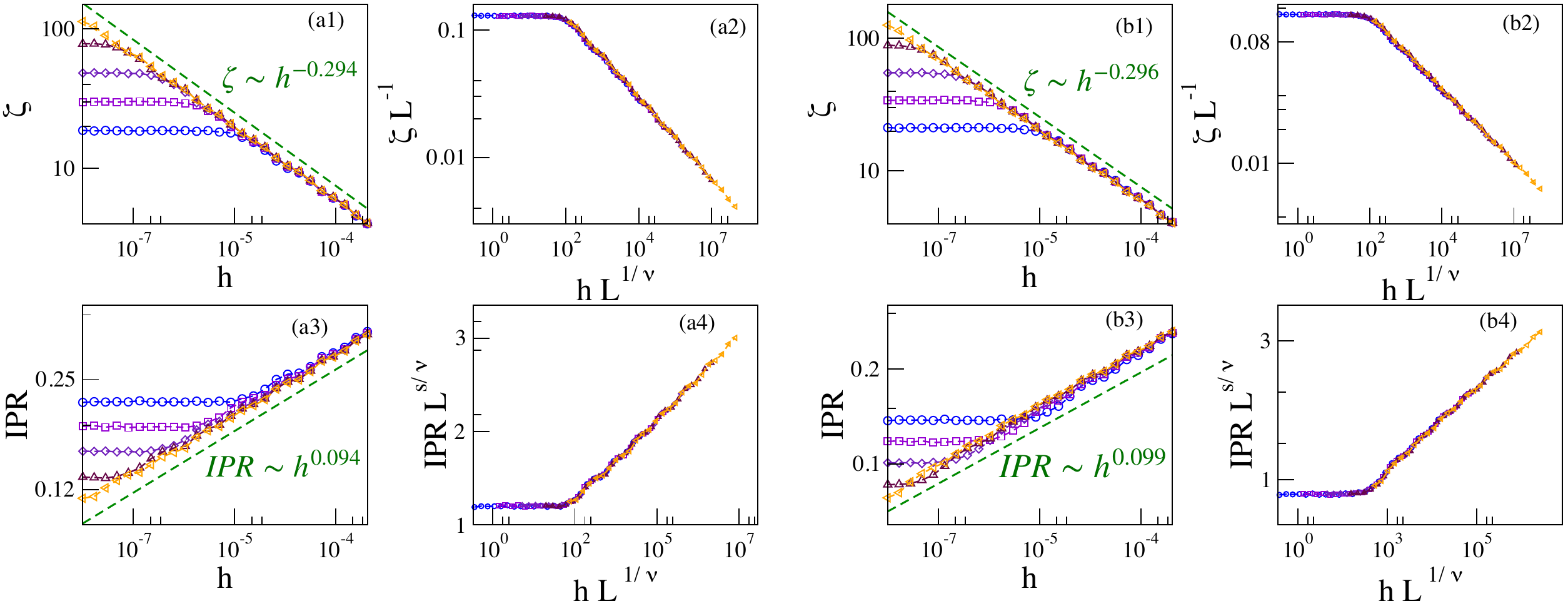}
\caption{{\bf Scaling Analysis for $\delta L^{1/\nu}=1$:}  {\bf (a1)} presents localization length $\zeta$ against the Stark field of strength $h$ for different system sizes, $L=144$ (blue circle), $L=233$ (violet square), $L=377$ (indigo diamond), $L=610$ (maroon triangle up), $L=987$ (orange triangle left). {\bf (a2)} shows the corresponding collapse plot for a value of the scaling exponent $\nu=0.29$. {\bf (a3)} displays the IPR versus $h$ for various values of $L$ which is aforementioned in (a1). {\bf (a4)} demonstrates the collapse plot of IPR using the scaling exponent value $s=0.096$. {\bf Scaling Analysis for $\delta L^{1/\nu}=-1$:} {\bf (b1)} presents localization length $\zeta$ against stark potential strength $h$ for different system sizes, $L$ with the same line-point scheme as before. {\bf (b2)} shows the collapse plot for a value of scaling exponent $\nu=0.29$. {\bf (b3)} displays the IPR versus $h$ for various values of $L$ which is aforementioned in (a1). {\bf (b4)} demonstrates the collapse plot of IPR using the scaling exponent value $s=0.096$.}
    \label{fig:fig2}
\end{figure*}

We study the dependence of the localization length  $\zeta$, of the ground state, on the control parameter $h$ for different system sizes $L$. For localization length, we adopt the following scaling ansatz,
\begin{eqnarray}
\label{eq:zeta_collapse_function}
\zeta/L = f_1(h L^{1/\nu})
\end{eqnarray}
where $f_1[\cdot]$ is an arbitrary function. This scaling ansatz is confirmed through the methodology of data collapse. The concept behind data collapse is as follows: We collect the data $\zeta / L$ versus $h L^{1/\nu}$ for various system sizes $L$, with a specific value of $\nu$, for which all the datasets collapse perfectly. Initially, by some trial and error, we manually find out a rough range of the scaling exponent for which the data collapse is quite well. Furthermore, for a better estimation of the scaling exponent we employ the cost function approach (see Appendix A for details). 

Given that the details of $\zeta$ depend on a particular choice of $\phi$, a configuration averaging over 5000 random realizations of $\phi$ is performed to eliminate the $\phi$ dependent fluctuations. In the single-particle case, apart from $\zeta$, similar averaging is performed for all other quantities of our interest, to be introduced later. In Fig.~\ref{fig:fig1}(a1), we depicted $\zeta$ against $h$ for various values of $L$. This figure illustrates that for finite-sized systems, an initial flat region exists wherein the corresponding wave function exhibits an extended nature, indicating that the ground state of the Hamiltonian for $\delta = 0$ is equally probable across this phase. Beyond a certain threshold of $h$, the localization length exhibits weak dependence on the system size, leading to the localization of the wave function. 

Setting $|g-g_c|=h$ in Eq.~\eqref{eq:zeta},  one finds the localization length exponent to follow the relation $\zeta \propto h^{-\nu}$, where the scaling exponent $\nu$ controls the rate of divergence of the localization length of the system in the thermodynamic limit near criticality. The scaling exponent, $\nu$, can be computed via two explicit approaches. A straightforward approach is to perform a simple fitting analysis on the finite-size data set. This is what is shown via the green dotted line in Fig.~2(a1). The best fit suggests $\nu = 0.296(2)$, where $(.)$ is the fitting error on the last significant digit reported in this work.  Hence, we obtain $\nu$ within $0.6\%$ uncertainty via best-fit. The data collapse method provides a second route for extracting $\nu$. Here the scaled localization length, $\zeta/L$, is plotted with the scaled Stark field, $h L^{1/\nu}$. Finite size data collapse can be obtained by tuning $\nu$ (see Eq.~\eqref{eq:zeta_collapse_function}). The quality of collapse is then examined via computation of the cost function, $C_Q$  (see Appendix A). The best collapse corresponds to the particular $\nu$, for which $C_Q$ is the minimum. The cost function approach further suggests the scaling exponent $\nu$ within a window $(\Delta \nu)$ between $0.286$ to $0.298$. $C_Q$ assumes minimal value within this window and remains flat throughout (see Appendix A). We report the optimal value of $\nu$ by taking an average over the flat window of $\Delta\nu$ and the uncertainty associated with the estimated value of the quantity is $\pm \Delta \nu / 2$. Hence, the optimal $\nu$ corresponding to the best collapse turns out to be $0.292(6)$. We demonstrate a collapse plot in Fig.~\ref{fig:fig1}(a2) utilizing the function specified previously in Eq.~\eqref{eq:zeta_collapse_function} for $\nu = 0.292$. Evidently, the reported values of $\nu$ from two different methods agree within the uncertainty range. We perform similar error analysis for all the cases to be discussed in the following.

The next observable that can capture the delocalization-localization transition is the IPR, defined as,
\begin{eqnarray}
\label{eq:IPR_of_wave_function}
\mathrm{IPR} = \sum_{i=1}^{L}p_i^2.
\end{eqnarray}
Near criticality, in the thermodynamic limit, the IPR scales as,
\begin{eqnarray}
\label{eq:IPR_Scaling_h}
\mathrm{IPR} \propto |g-g_c | ^{s},
\end{eqnarray}
where $s$ is another scaling exponent. For the pure AA model, $|g-g_c|=\delta$ and $s\approx0.33$; for the pure Stark model $|g-g_c|=h$ and $s\approx0.33$. On the other hand, near criticality, IPR scales with system size $L$ as, 
\begin{eqnarray}
\label{eq:IPR_Scaling_L}
\mathrm{IPR} \propto L^{-s/\nu}.
\end{eqnarray}
So for IPR, the scaling ansatz we use is,
\begin{eqnarray}
\label{eq:IPR_collapse_function}
\mathrm{IPR} = L^{- s/\nu}f_2(h L^{1/\nu}),
\end{eqnarray}
where $f_2[\cdot]$ is another arbitrary function.

The variation in the value of IPR with the increase in $h$ for different system sizes $L$ is depicted in Fig.~\ref{fig:fig1}(b1). In this figure, up to a certain region of $h$, the IPR values appear flat, indicating the delocalized nature of the wavefunction. Beyond this region, the IPR becomes independent of the system size and hence suggests the system's entrance into the localized phase. Considering that the IPR scales with $h$ as, $\text{IPR} \sim h^{s}$,  green-colored dashed line has corresponds to the best fit. The obtained value of $s$ from the best fit is given by, $s=0.097(2)$. Moreover, we again perform the cost function analysis for IPR as well. It suggests the value of $s$ as, $s = 0.0968(1)$. Fig.~\ref{fig:fig1}(b2) presents the best collapse plot of the IPR for $s = 0.0968$. Noticeably, in this case, the ratio $s/\nu$ $(s/\nu\approx0.33)$ coincides with the one from the pure AA model. Hence, the scaling of IPR with the system size is basically the same for both of these two models.\\
\begin{figure}[t]
    \centering
    \includegraphics[width=0.48\textwidth]{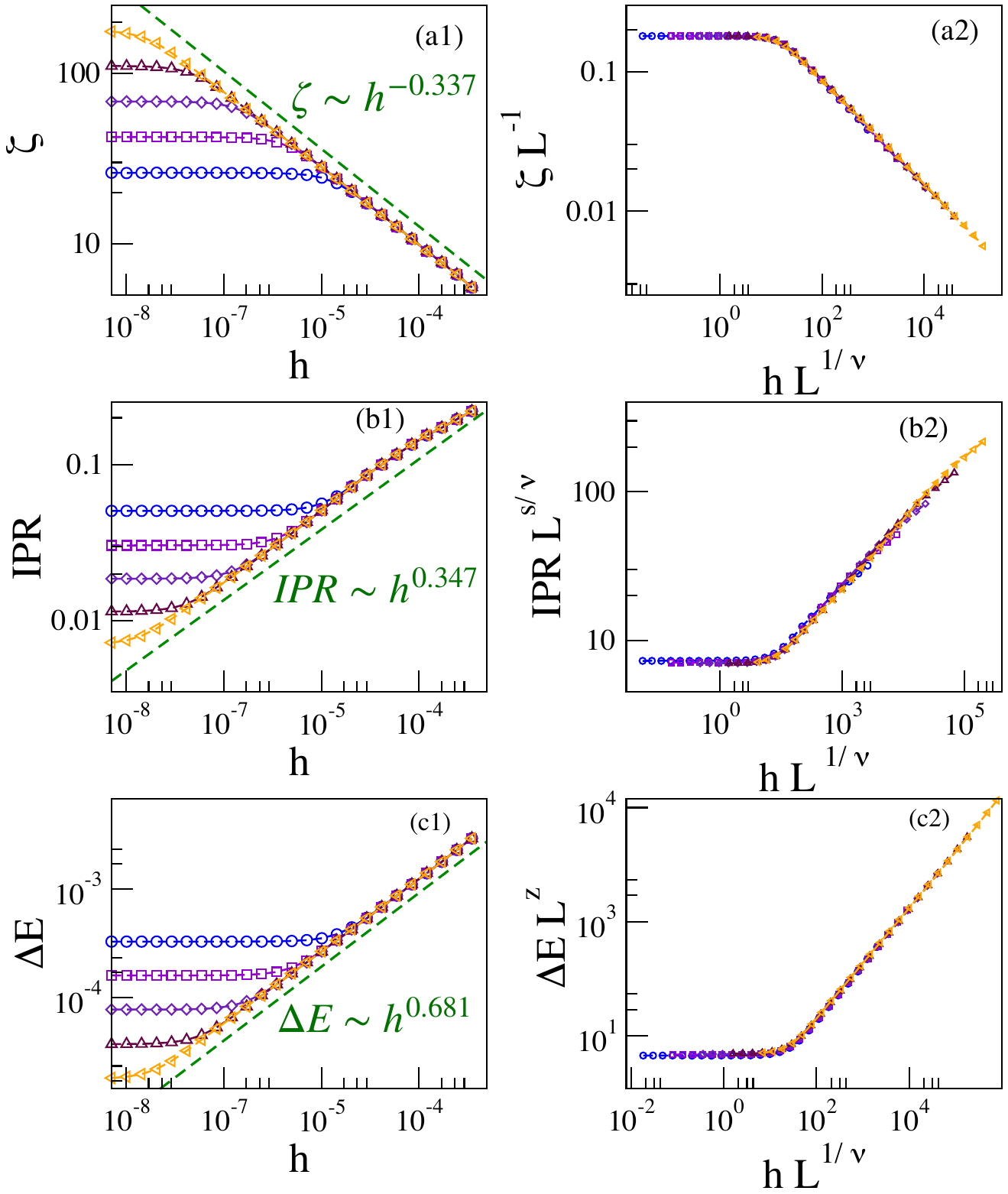}
\caption{{\bf Scaling Analysis for $\delta=-0.1$:} {\bf (a1)} presents localization length $\zeta$ against stark potential strength $h$  for different system size $L$, with the same line-point scheme. {\bf (a2)} shows the collapse plot for a value of the scaling exponent $\nu=0.33$. {\bf (b1)} displays the IPR versus $h$ for various values of $L$. {\bf (b2)} demonstrates the collapse plot of IPR with $s=0.33$. {\bf (c1)} depicts the variation of $\Delta E$ as a function of $h$ for different $L$'s. {\bf (c2)} demonstrates the collapse plot of $\Delta E$ using the scaling exponent $z=2$ as before.}
    \label{fig:fig3}
\end{figure}

Finally, to elucidate the localization transition, we examine the energy gap between the ground state and the first excited energy, denoted as $\Delta E$. In the thermodynamic limit, the energy gap scales with the control parameter as, 
\begin{eqnarray}
\label{eq:energy}
\Delta E \propto |g-g_c|^{\nu z}.
\end{eqnarray}
For the pure AA model $z=2.37$ and for the pure Stark model $z = 2$. To perform the finite size scaling for $\Delta E$, we use the scaling ansatz:
\begin{eqnarray}
\label{eq:energy_collapse_function}
\Delta E = L^{-z}f_3(h L^{1/\nu}),
\end{eqnarray}
where $f_3[\cdot]$ is another arbitrary function.

Fig.~\ref{fig:fig1}(c1) shows, the energy gap $\Delta E$ against the Stark field strength $h$ for different system sizes. In this case, also there is an initial flat region showing a very narrow energy gap in the delocalized phase. Beyond a certain strength of Stark field $h$, $\Delta E$ becomes independent of the system size, and the energy gap gets increased with an increase in the field strength. From the fitting function $\Delta E \sim h^{\nu z}$, as depicted by the green-colored dashed line in Fig.~\ref{fig:fig1}(c1), We obtain, $\nu z = 0.702(1)$. The value of $z$ can be derived by simply dividing $\nu z$ by $\nu$.  We find $z = 2.371(1)$. The cost function approach via data collapse also suggests that $z = 2.371(1)$. Similar to the scaling exponent associated with IPR, the scaling exponent $z$ again remains the same as the pure AA model. The collapse plot corresponding to $\Delta E$ is presented in Fig.~\ref{fig:fig1}(c2).

\section {Scaling Properties near AA criticality}
In the previous section, we have done the scaling analysis at the AA criticality, i.e., $\delta=0$. Now we turn our attention to the near criticality region of the AA model. Now as $\delta\ne0$, one can expect that the previous scaling ansatz will not work properly and some modifications are needed. 

We start with the scaling of localization length. In the presence of both $\delta$ and $h$, we use the following ansatz for the scaling of localization length,
\begin{eqnarray}
\label{eq:zeta_collapse_hybrid}
\zeta/L = f_4(\delta L^{1/\nu_\delta}, h L^{1/\nu}),
\end{eqnarray}
where $\nu_\delta$ is the scaling exponent for localization length in the case of a pure AA model. The novelty of this ansatz is that when $\delta=0$, it coincides with the ansatz given in Eq.~(\ref{eq:zeta_collapse_function}), and for $h=0$, it becomes the scaling ansatz for pure AA model, i.e.,   $\mathrm{\zeta}\propto\delta^{-\nu_\delta}$, in the thermodynamic limit.

In the same spirit, we define the scaling ansatz for IPR as,
\begin{eqnarray}
\label{eq:IPR_collapse_hybrid}
\mathrm{IPR} = L^{-s_\delta/\nu_\delta} f_5(\delta L^{1/\nu_\delta}, h L^{1/\nu})
\end{eqnarray}
where $s_\delta$ is the scaling exponent of IPR in the pure AA model ($h=0$), i.e.,   $\mathrm{IPR}\propto\delta^{s_\delta}$, in the thermodynamic limit. When $\delta=0$,  this scaling ansatz takes the similar form used in the previous section for the collapse as we have found that the $s/\nu$ value is the same for both the pure AA model and our model under consideration. Similarly, for $\Delta E$ it will take the form,
\begin{eqnarray}
\label{eq:delE_collapse_hybrid}
\Delta E = L^{-z_\delta} f_6(\delta L^{1/\nu_\delta}, h L^{1/\nu}),
\end{eqnarray}
where $z_\delta$ appears in the scaling exponent of $\Delta E$ in the pure AA model, i.e., $\Delta E\propto\delta^{\nu_\delta z_\delta}$, in the thermodynamic limit, again this works due to the same value of $z$ in both models.

To examine the validation of the above scaling ansatz, we now turn our attention to delve into scaling properties of $\zeta$, IPR, and $\Delta E$ near criticality for $\delta L^{1/\nu}=1$ and $\delta L^{1/\nu} = -1$. Upon taking $\delta$ in this fashion for different $L$, we expect that our previous critical exponents would work as we are still in the vicinity of the AA critical point. In Fig.~\ref{fig:fig2} both these cases are shown and we have numerically found that the collapse still happens for the same set of scaling exponents as found in the case of $\delta=0$. The value of $\nu$ obtained from fitting for $\delta L^{1/\nu} = 1$ is $0.295(4)$. For $\delta L^{1/\nu} = -1$, the value of $\nu$ is $0.297(2)$. For $\delta L^{1/\nu} = 1$, the value of $s$ is $0.095(2)$, while for $\delta L^{1/\nu} = -1$, the value of $s$ is $0.099(2)$. Based on the cost function approach for $\delta L^{1/\nu} = 1$, the value of $\nu$ lies in the range $0.285$ to $0.298$, giving an average of $0.292(1)$. The corresponding value of $s$ obtained is $s = 0.0964(1)$. Similarly, for $\delta L^{1/\nu} = -1$, the value of $\nu$ is in the range $0.285$ to $0.298$, with an average of $0.291(5)$, and the value of $s$ obtained is $s = 0.0972(1)$.

\begin{figure}[t]
    \centering
\includegraphics[width=0.48\textwidth]{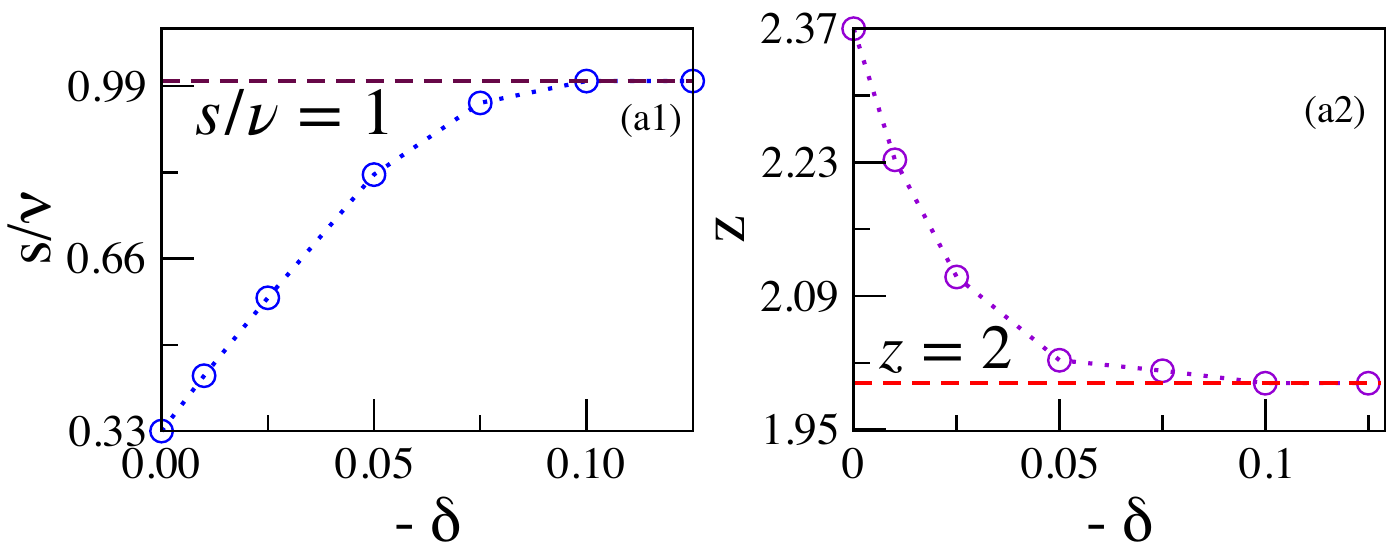}
    \caption{{\bf (a1)} presents $s/\nu$ for different values of $\delta$. For $\delta=0$ the $s/\nu$ value is $0.33$ which is the same as that of the pure AA model and this value continuously increases to $1$, which is the same as the pure Stark model, as one moves away from the AA criticality with negative $\delta$ values. {\bf (a2)} presents the plot of $z$ for different values of $\delta$. In this case, there is a continuous change in the value of $s/\nu$ from 2.37, which is the same as the pure AA model, to 2, which is for the pure Stark model, as one moves away from the criticality.}
    \label{fig:fig4}
 \end{figure}

 \begin{figure}[t]
    \centering
    \includegraphics[width=0.48\textwidth]{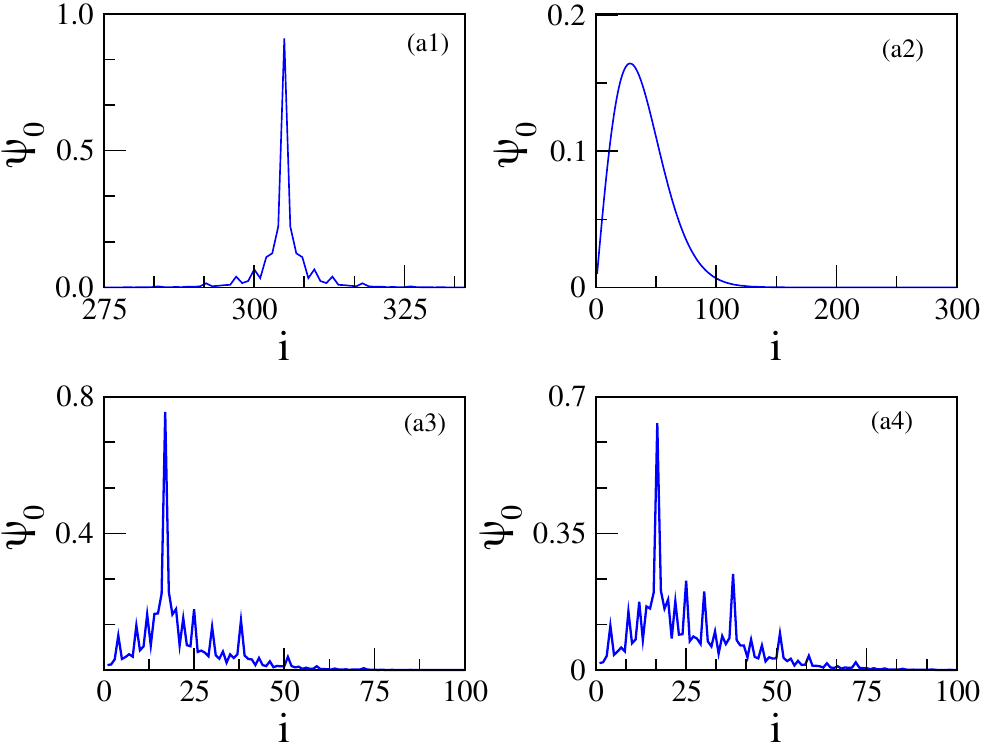}
    \caption{{\bf Ground State Wavefunction} $\boldsymbol{\psi_0:}$  {\bf (a1)} describes the ground state wavefunction for the pure AA model, where $h=0$ and $\delta= 0.4J$. {\bf (a2)} presents the ground state wavefunction for the pure stark model, where $\delta=-2J$ and $h=10^{-4}$. {\bf (a3)} presents the ground state wavefunction for $\delta = 0$ with $h=10^{-4}$ and {\bf (a4)} presents  the wave function for $\delta = -0.1$ and $h=10^{-4}$. All plots are done for system size 610 and for the AA potential, the phase, $\phi=0$.}
    \label{fig:fig5}
\end{figure}

\begin{figure}[t]
    \centering
\includegraphics[width=0.45\textwidth]{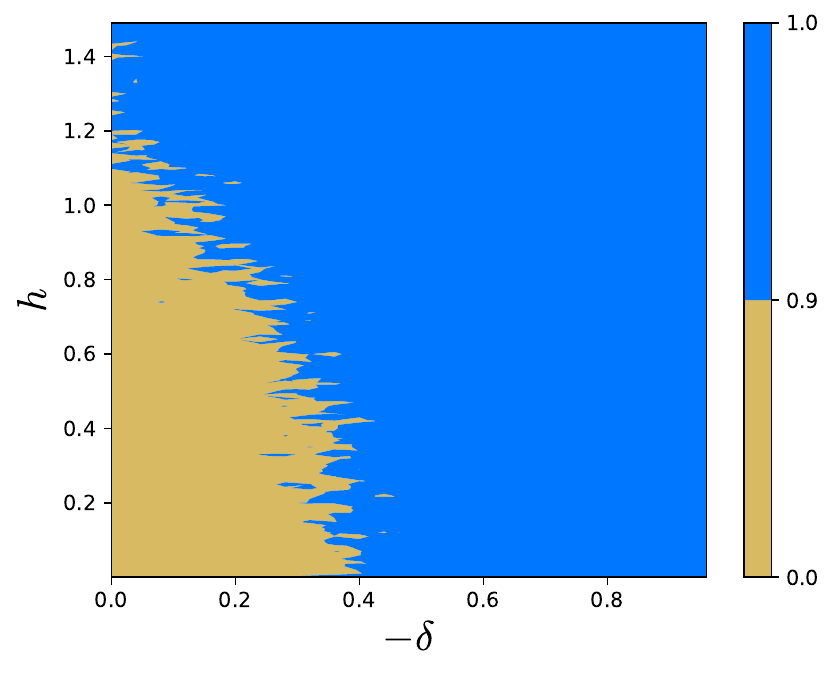}
\caption{{\bf } We show the fidelity $\mathcal{F}(h,\delta)$ between the ground state of the AAS system (i.e., the system under the dual potential), say $|\psi_0(h,\delta)\rangle$, with the ground state of the pure stark model, i.e., $|\psi_0(h,\delta=0)\rangle$, in the parametric space of $h$ and $\delta$ for system size $L=610$. The range of the quantitative value is shown in the color bar. We have performed a configuration averaging over $\phi$ with 100 random samples. The blue region corresponds to the parametric space with high fidelity (90\% or more) indicating negligible influence of the AA potential, and hence, implies Stark dominated localization. } 
 \label{fig:schamatic_appendix}
\end{figure}

\section{Scaling Analysis at Far from the Aubry-Andr\'e Criticality and Stark dominated localization}

Now we focus on the scaling analysis at comparatively far away from the AA criticality. Far from the AA criticality, the wavefunction is supposed to be delocalized as the effect of AA potential is faint at that region. In order to validate this understanding we perform a scaling analysis for a fixed value of $\delta < 0$, specifically $\delta = -0.1$, as illustrated in Fig.~\ref{fig:fig3}.

In Fig. \ref{fig:fig3}(a1), the localization length ($\zeta$) is plotted against $h$, while in Fig. \ref{fig:fig3}(a2), we depict $\zeta / L$ against $h L^{1/\nu}$, with a specific value of $\nu = 0.33$. Remarkably, all datasets collapse onto a single curve for this value of $\nu$. From the fitting function in \ref{fig:fig3}(a1), $\zeta \sim h^{-\nu}$, We get $\nu = 0.337(5)$. This observation suggests that as we move away from the AA criticality the scaling nature replicates the scaling of Stark localization. The IPR for $\delta = -0.1$ across various system sizes is shown in Fig. \ref{fig:fig3}(b1). Scaling analysis for IPR gives the value, $s = 0.347(3)$, which is confirmed by data collapse shown in Fig. \ref{fig:fig3}(b2) and also by the cost-function approach. This scaling exponent is identical to that of the pure Stark model. In Fig. \ref{fig:fig3}(c1), we plot the dependence of $\Delta E$ on $h$. The nature of the dependence is found to be the same as before. The collapse plot corresponding to $\Delta E$ is depicted in Fig.~\ref{fig:fig3}(c2), yielding a critical exponent of $z = 2.00(2)$ which is again the same critical exponent value for that of the pure Stark model. In Fig. \ref{fig:fig3}(c1) from the scaling of $\Delta E$ with $h$, the green color fitting function gives the value $\nu z = 0.682(2)$ with $0.32\%$ error and the value of $z$ turns out to be $1.989(1)$. 
From the cost function, the range of $\nu$, for which $C_Q$ remains minimal and flat is from $0.326$ to $0.344$. The average of $\nu$ within this window is $0.335$. Finally, the extracted values of $s$ and $z$ are $0.03457(1)$ and $1.997(1)$, respectively.

Thus, as we move away from the AA criticality the system does not experience much effect due to the AA potential. Likewise, the delocalization-localization transition is dominated by the Stark field. Therefore we study the behaviour of the scaling exponents as we slowly move away from the criticality. Fig.~\ref{fig:fig4} nicely depicts this variation. The variation of $s/\nu$ and $z$ are shown in  Fig.~\ref{fig:fig4}(a1) and Fig.~\ref{fig:fig4}(a2) respectively. To be specific, here, we only focus on the negative values of $\delta$, as a slight departure from the critical point in the positive $\delta$ values results in an unusual variation in the characteristic observables due to the finite localization of the underlying AAH potential, and as a consequence, the scaling analysis fails in this region. This happens due to the competition between the two kinds of localization. From Fig.~\ref{fig:fig4} it is easily noticeable that in the negative $\delta$ region there is a continuous change in the values of the scaling exponents. As we move away from the AA criticality, the values of $s/\nu$ and $z$ slowly change from that of the pure AA model to the pure Stark model. 

Moreover, in order to understand the competing nature of two distinct types of localization-inducing potentials closely, it is interesting to directly look into the structures of the ground-state wavefunctions for various cases. We present some exemplary cases in Fig. \ref{fig:fig5} for various scenarios relevant to our study of the scaling nature of the system near the criticality. The reflection symmetry of the system is strongly broken in the presence of the Stark potential, and, correspondingly, the ground-state localized wavefunction in the pure Stark of the system is strongly peaked near one of the ends of the lattice, unlike the pure AA case. Setting the AA potential near the AA criticality introduces exotic structures characterized by small peaks around a central large peak, because of the strong influence of the AA potential. As $|\delta|$ increases, i.e., for which the pure AA system is delocalized, and as $h$ increases, the effects of the AA potential in the AAS system gradually diminish, as one may expect. In order to explain this, which has also been presented in the schematic diagram Fig.~\ref{fig:schamatic}, we compute the quantity, $\mathcal{F}(h,\delta) = |\langle \psi_0(h,\delta)|\psi_0(h)\rangle|$, which is the fidelity between the ground state of the system of AAS model $|\psi_0(h,\delta)\rangle$ and that of the pure Stark model $|\psi_0(h,\delta=0)\rangle$ and provides an estimate on the closeness of these two quantum states. The high value of $\mathcal{F}$ essentially implies that the underlying physics of the AAS model is effectively governed by the Stark potential. In Fig.~\ref{fig:schamatic_appendix}, We have shown the bicolor plot, whose x axis is $-\delta$, the y axis is $h$ and the colored region is described by $\mathcal{F}(h,\delta)$. The blue region is where $\mathcal{F}(h,\delta)\ge 0.9$. The localization in this region is essentially stark dominated. The yellow region is for $0\le\mathcal{F}(h,\delta)\le 0.9$. The AA potential has a strong influence near $|\delta| \to 0$ and $h=0$, which gradually decreases with increasing values of $h$ and $\delta$. As discussed in the previous paragraph, in the limit of $h \to 0$, the manifestation of Stark dominated has been demonstrated by tracking the evolution of the scaling exponents with the change of $|\delta|$. In addition, the system is sufficiently away from the AA criticality ($|\delta|=0.1$), the scaling exponents turn out to be identical to the pure Stark system. As one may expect, $\mathcal{F}(h,\delta)$ is quite high for $|\delta| \ge 0.1 ( \mathcal{F} \gtrsim 0.84\%)$.

\begin{figure}[t]
    \centering
    \includegraphics[width=0.48\textwidth]{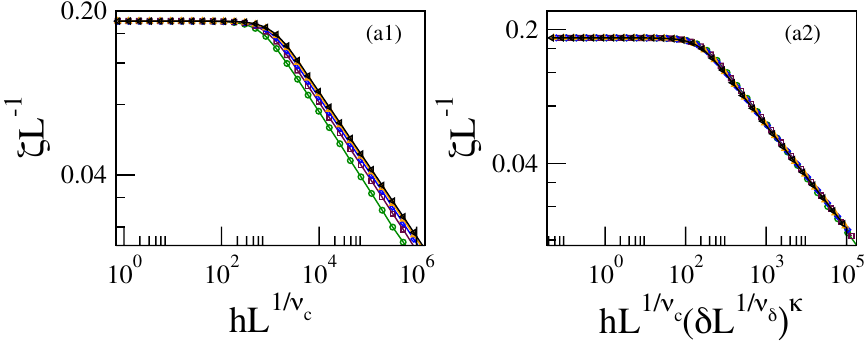}
    \caption{In {\bf (a1)}, we have plotted the curves of $\zeta L^{-1}$ against $h L^{1/\nu_c}$ for fixed system size, $L=377$, having various $\delta$ values: $\delta=-0.1$ (green circle), $\delta=-0.2$ (maroon square), $\delta=-0.3$ (blue diamond), $\delta=-0.4$ (orange triangle up), and $\delta=-0.5$ (black triangle left). Here $\nu_c = 0.29$, the value of scaling exponent for localization length at the AA criticality, i.e., $\delta=0$. {\bf (a2)} shows the collapse plot of $\zeta L^{-1}$ versus $hL^{1/\nu_c} (\delta L^{1/\nu_\delta})^{\kappa}$. Here also, we have taken $\phi$ average for 5000 choices. }
    \label{fig:fig6}
 \end{figure}

\vspace{0.5cm}
\section{Hybrid Scaling Exponent}
Next, we focus on a case, where the system size, $L$, is kept fixed and different $\delta$ values are considered with $\delta<0$. For this case, we use the following scaling ansatz,
\begin{eqnarray}
\label{eq:delE_collapse_hybrid}
\zeta/L = f_7(h L^{1/\nu_c} (\delta L^{1/\nu_\delta})^{\kappa}),
\end{eqnarray}
where $\nu_c$ is the scaling exponent for the localization length of the ground state in our model, at the AA criticality, i.e., $\delta=0$. Here we numerically find out the new scaling exponent $\kappa$. Near AA criticality, this $\kappa$ should have the following form,
\begin{eqnarray}
\label{eq:delE_collapse_hybrid}
\kappa = \nu_\delta (1/\nu_s-1/\nu_c),
\end{eqnarray}
where $\nu_\delta$ and $\nu_s$ are the critical exponents for the localization length in the case of pure AA and Stark model.

In Fig.~\ref{fig:fig6}(a1), we show the variation of $\zeta/L$ against $hL^{1/\nu_c}$ for different values of $\delta$. Fig.~\ref{fig:fig6}(a2) shows that the collapse plot of $\zeta L^{-1}$ as a function of the hybrid quantity $hL^{1/\nu_c} (\delta L^{1/\nu_\delta})^{\kappa}$. From the cost function approach, we have found the scaling exponent $\kappa=-0.418(2)$ which matches with the value obtained via Eq.~\ref{eq:delE_collapse_hybrid}.

\section {Application as quantum sensor} 
Quantum sensing, a crucial pillar of quantum technology, enables the estimation of unknown parameters, e.g., magnetic fields, electric fields, gravitational fields, time, and other physical quantities, with remarkable precision. Recent experimental advancements towards developing quantum sensing devices, which include interferometry \cite{interferometry, Nagata07}, magnetometry \cite{Wasilewski10, Sewell12}, and ultracold spectroscopy \cite{Liebfried04, Roos06}, among others, demonstrate the advantages originating from the quantum domain over the conventional classical counterparts. Recently, several works have proposed criticality-based adiabatic quantum sensing devices. The motivation behind leveraging criticality for developing quantum sensors lies in the fact that a many-body system is susceptible to small changes in a system parameter near criticality \cite{Zanardi08}. In the context of measuring an unknown field, it has been demonstrated that QFI, $F_Q$ (see below for a formal definition), can saturate the Heisenberg limit (HL) at the localization-delocalization transition in the pure AA  model, implying a quantum advantage in precision measurement \cite{Sahoo24}. Similarly, localization transition for pure stark model enables $F_Q$ to beat HL \cite{He23}, and hence, paves the way towards designing an efficient quantum weak-field sensor. Here, our goal is to show that a better quantum sensor can be engineered by introducing an additional control field in the form of the AA potential for precision estimation of the Stark weak field. In the following, we briefly discuss the idea of quantum sensing and present the formal definition of the QFI, $F_Q$.

One needs to perform measurements for the estimation of an unknown parameter $h$. Let's consider that the parameter $h$ is encoded in a quantum state $\rho(h)$. In the projective measurement basis $\{\pi_n$\}, the probability corresponding to the $n^\text{{th}}$ outcome is given by, $p_n(h) = \text{Tr}[\rho(h) \pi_n]$. In this situation, the standard deviation of $h$ is bounded by the relation $\delta h \geq 1/\sqrt{M F_c}$ \cite{Cramer46, Helstrom76}, where $M$ is the number of measurements, and the Classical Fisher information, $F_c$ is defined as,
\begin{equation}
    F_c = \sum_n \frac{1}{p_n(h)} \left(\frac{dp_n(h)}{dh}\right)^2,
\end{equation}

\begin{figure}[t]
\centering
\includegraphics[width=0.45\textwidth]{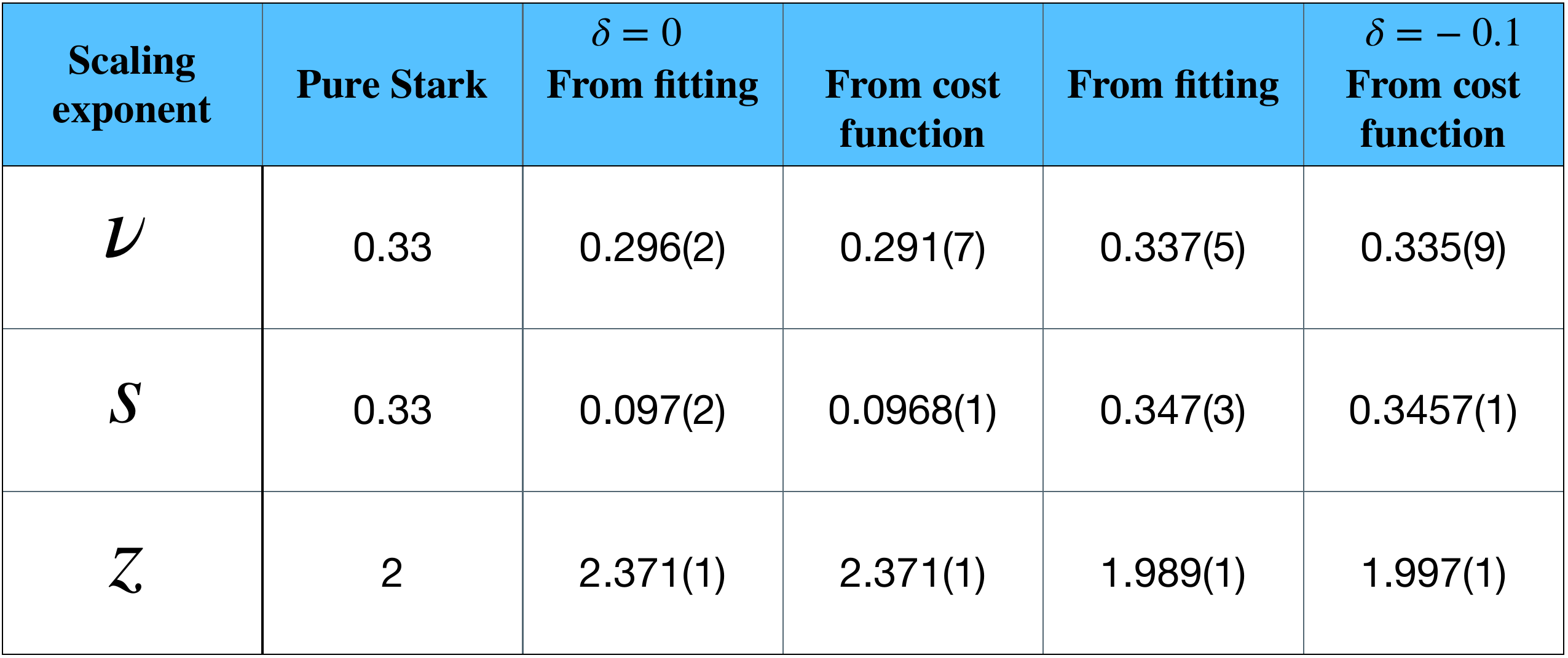}
 \caption{The scaling exponents $\nu$, $s$ and $z$ at and near the AAH criticality are tabulated.}
    \label{fig:table}
\end{figure}

\begin{figure}[t]
    \centering
\includegraphics[width=0.4\textwidth]{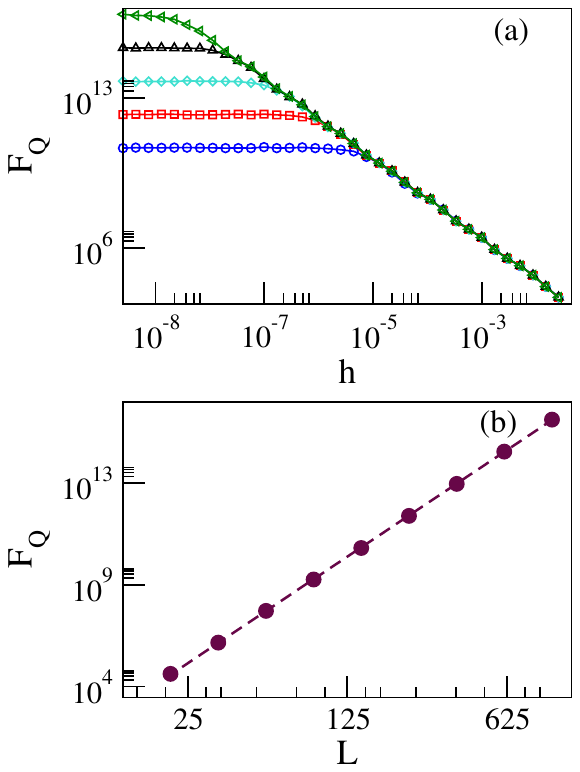}
\caption {{\bf Single party QFI:} {\bf (a)} presents  QFI against the stark field $h$ for system size L = 144 (blue circle), 233 (red square), 377 (turquoise diamond), 610 (black triangular up), 987 (green triangular left).  {\bf (b)} shows the scaling plot. The maroon circles are QFI for different system sizes $L = 21, 34, 55, 89, 144, 233, 377, 610$ at $h=10^{-9}$, and the dotted line is the best fit $F_{Q} \sim L^{6.7}$. Both plots are shown in log-log scale. A configuration averaging over $\phi$ is performed with 8000 random samples.} 
 \label{fig:sensing}
\end{figure}

Maximization over all possible measurements, one can obtain a measurement-independent quantity, the so-called  QFI, $F_Q$. The QFI can be expressed as, $F_Q = \text{Tr}[L_h^2 \rho(h)]$, where $L_h$ is the symmetric
logarithmic derivative that is defined implicitly as $ \partial_h \rho(h) = 1/2 \{\rho(h),L_h\}$. For a pure state, $\rho(h) =  |\psi (h) \rangle \langle \psi (h)|$, the $L_h$ can be written as $2\partial_h \rho(h)$. The QFI now turns out to be, 
 \begin{equation}
     F_Q = 4  (\langle \partial_h \psi(h)|\partial_h \psi(h)\rangle - |\langle \partial_h \psi(h)|\psi(h)\rangle|^2 ).
     \label{QFI-eq}
 \end{equation}
 In the many-body context, system size is a resource, and hence, the scaling of QFI, $F_Q$, with system size, $L$, is of key interest \cite{Zanardi08}. 
 
 Considering QFI scales as, $F_Q \sim L^{\beta}$, $\beta=1$ corresponds to the standard quantum limit (SQL), which is what a classical system can achieve at best with $L$ probes. A definitive quantum advantage corresponds to a case, $\beta > 1$. The special case of $\beta = 2$ is the so-called Heisenberg scaling, beyond which, i.e., for $\beta > 2$, is termed as super-Heisenberg scaling.  Recently, it has been shown that the Stark transition can be exploited for precision measurement of an unknown weak field \cite{He23}.  There the ground state QFI scales as $F_Q \sim L^{5.9}$, and hence, achieves a super-Heisenberg scaling with the system size. In the following, as mentioned before, we discuss whether it is possible to obtain an enhanced scaling of the QFI in the presence of an additional AAH potential that we consider as a control field, and thus it is possible to design a better quantum sensor for weak field estimation.

We consider the control field, i.e., the AA potential is set at AA criticality, i.e., $\delta = 0$. Considering the ground state as the probe state, we compute QFI using Eq.~(\ref{QFI-eq}). In Fig.~\ref{fig:sensing}, we present $F_Q$  as a function of the stark strength, $h$, for various system sizes $L= 144, 233, 377, 610, 987$. Throughout the calculations, averaging over  $\phi$ has been performed with OBC. The figure illustrates that, for finite systems, as the Stark strength $h$ gradually increases, the QFI remains flat in the extended region. However, beyond this flat region, the QFI decreases in the localized region. Evidently, in the flat areas of the finite systems, the QFI has a finite-size scaling, whereas it becomes scale-invariant in the localized phase, as one may expect. Fig.~\ref{fig:sensing}(b) illustrates the scaling of QFI with system size in the extended region. The maroon circles represent QFI values for different system sizes, and the maroon dashed line shows the fitting function $F_Q \sim L^{\beta}$, where $\beta$ is 6.70(1).  Evidently, the
$F_Q$ at the transition point scales as $F_Q \sim L^\beta$, where $\beta$ turns out to be $\beta = 2/(d \nu)$ with 
$d$ as the spatial dimension.  The scaling exponent $\nu \sim 0.29$ is associated with the localization length of the hybrid system under the influence of two different kinds of localizing potentials (see Sec.~III). It's worth mentioning here that, in the context of the second-order phase transition, the QFI is characterized by  $\nu$ -- the scaling exponent associated with the localization length. In the localization transition induced many-body quantum sensors, the scaling exponent associated with the localization length plays the same role in the QFI. The enhanced scaling of QFI in the presence of the AA criticality establishes the superiority of the system under consideration as a many-body quantum sensor for precision measurement of the weak field in comparison to the pure Stark sensors.

\begin{figure*}[t]
\centering
\includegraphics[width=0.94\textwidth]{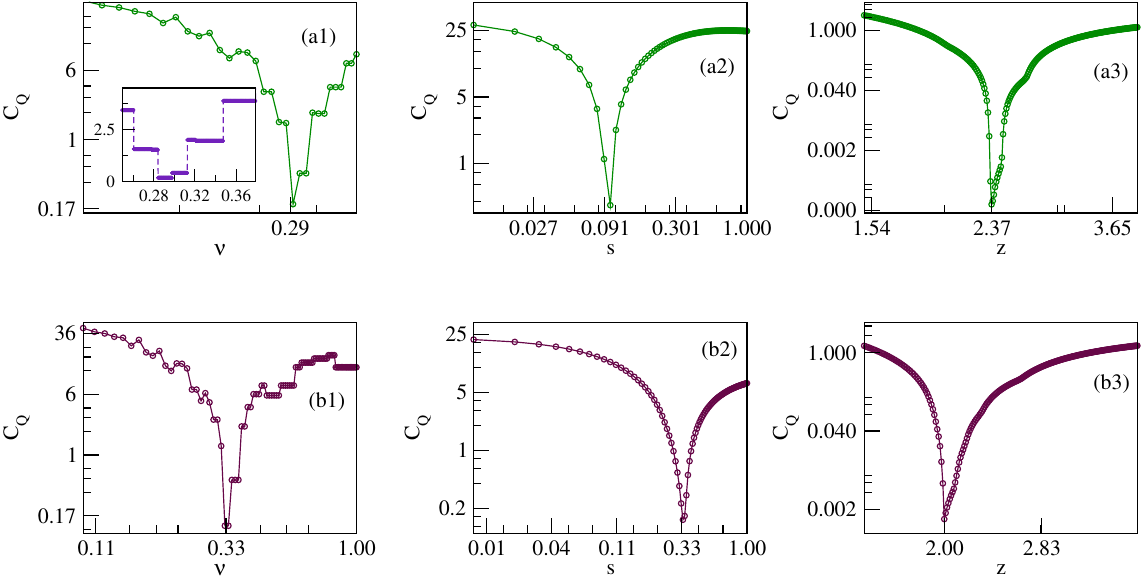}
 \caption{{\bf Cost function for $\delta = 0$:} {\bf (a1)} presents plot for cost function with respect to $\nu$. It shows that the $C_Q$ has global minima for $\nu = 0.29$. The inset shows an enlargement near the global minima. {\bf (a2)} shows plot $C_Q$ versus $s$ and the minimum value of $C_Q$ is for $s=0.096$. {\bf (a3)} demonstrates the plot $C_Q$ versus $z$.  The minimum value of $C_Q$ is at $z = 2.37$. {\bf Cost function for $\delta = -0.1$:} {\bf (b1)} presents $C_Q$ versus $\nu$. The $C_Q$ has the lowest value for $\nu = 0.33$. {\bf (b2)} shows the plot for $C_Q$ versus $s$. Here also at $s = 0.33$, the $C_Q$ has global minima. {\bf (b3)} displays the $C_Q$ versus $z$ and the minimum value of $C_Q$ is for $z = 2$.}
\label{fig:cost}
\end{figure*}

\section {CONCLUSION} In this work we have explored the Stark localization in the presence of a quasi-periodic potential, namely the AA potential.  This exploration is based on the scaling analysis of finite-size systems. First, we have studied the effect of increasing Stark field at the AA criticality. We have found that the scaling exponents $\nu$ and $s$ are quite different than both the pure AA model and the pure Stark model. The value of $s/\nu$, however, is found to be the same as that of the pure AA model. Also, the scaling exponent $z$ is identical to that of the pure AA model. It indicates although the Stark field drives the transition, the transition is mostly AA type. Next, we have studied the same near the AA criticality. As in that case, we have used $\delta\ne0$, we cannot employ the scaling ansatz for one control parameter-based phase transition. We have developed some scaling ansatz that is capable of extracting the relevant scaling exponents quite efficiently. We tabulate the critical exponents at and near the AAH potential in Fig.~\ref{fig:table} as a summary. In addition, we consider a fixed system size and vary the distance from the AA criticality. For this, a hybrid scaling ansatz is used and it successfully describes the scaling nature. It is evident from our numerical results that critical exponent violates Harris-Chayes criterion \cite{Harris74,Chayes86} which suggests $\nu \ge 2/d = 2$, where $d$ is a spatial dimension.   In this respect it may be interesting to investigate the scope of a field theoretical treatment, e.g., via real space renormalization group theory, in providing an alternate route for the understanding of the localization transition in such systems under study. Finally, we investigate the variation of $s/\nu$ and $z$ with negative values of $\delta$, which reveals the fact that near the AA criticality, these exponents coincide with that of the pure AA model, and as we move away from the AA criticality we enter into the region where the localization is solely driven by Stark field with almost no trace of the presence of AA potential. We complement these studies with a fidelity based approach in order to gauge the parametric regime, where the underlying localization physics is essentially Stark dominated. These studies find an immediate application towards engineering an efficient Stark weak field sensor. We have shown that the presence of an AA control field results in an enhanced scaling of the QFI in comparison to the pure Stark model, establishing the hybrid system as a better sensing device. A similar idea may be useful in other contexts, in general, e.g., using additional control parameters for designing better criticality-based sensing devices across various kinds of quantum phase transitions.



\appendix
\renewcommand\thefigure{\thesection.\arabic{figure}}    
\setcounter{figure}{0}    
\section{Cost Function}
\label{costfn}

The scaling exponents are extracted via a cost function, $C_Q$, approach, where $C_Q$ is defined as \cite{Suntajs20},
\begin{eqnarray}
\label{eq:cost_function}
C_Q = \frac{\sum_{i=1}^{N_p-1} |Q_{i+1}-Q_i|}{\mathrm{max}\{Q_i\}-\mathrm{min}\{Q_i\}} -1,
\end{eqnarray}
where $\{Q_i\}$ is the dataset for different values of $h$ and $L$ and $N_p$ is the total number of data points in the dataset. Sorting all $N_p$ values of $|Q_i|$ according to the increasing value of $L\,\mathrm{sgn}[h-h_c]\,[h-h_c]^\nu$, the best scaling exponent can be found for the minimum value of cost function $C_Q$. Ideally, the perfect data collapse is found for $C_Q = 0$, i.e., $\sum_{i=1}^{N_p-1} |Q_{i+1}-Q_i| = \mathrm{max}\{Q_i\}-\mathrm{min}\{Q_i\}$. 

In the following, we present certain representative cases demonstrating the behavior of the cost function around the estimated critical exponents. We plot $C_Q$ with respect to the critical exponents $\nu$, $s$ and $z$ in Fig.~\ref{fig:cost}. The Fig.~\ref{fig:cost}(a1-a3) demonstrate the plot for $C_Q$ versus $\nu$, $s$ and $z$, respectively for $\delta = 0$. The figures depict that the $C_Q$ has a global minimum and the minimum value of $C_Q$ is at $\nu = 0.29$, $s = 0.096$, and $z=2.37$, these values agree the value of scaling exponents that extract from collapse plot. Similarly for $\delta = -0.1$, the $C_Q$ versus $\nu$, $s$, and $z$ are shown in Fig.~\ref{fig:cost}(b1-b3), respectively. Here also $C_Q$ has a local minimum. All data collapses for the minimum value of $C_Q$ at $\nu = 0.33$, $s = 0.33$, and $z=2$ for $\delta = -0.1$.


\end{document}